# Comment on **Proof that the Hydrogen-Antihydrogen Molecule is Unstable**

G. Van Hooydonk, Ghent University, Faculty of Sciences, Krijgslaan 281, B-9000 Ghent (Belgium)

**Abstract.** *The claim by Gridnev and Greiner that molecule* H$\underline{H}$ *is unstable cannot be a proof as it is based on a wrong conjecture. This is illustrated with 4 examples, including observed natural hydrogen-antihydrogen oscillations never detected previously.*

The conjecture that *with pure Coulomb forces no bound state of hydrogen-antihydrogen exists* [1] is not absolutely true [2-3], since *pure Coulomb forces* give stable H$\underline{H}$ [4-5]. Linking H$\underline{H}$-stability with (*Jacobi*) mass [1] is ambiguous, as particle mass is related with charge-separation. We give 4 *pure Coulomb effects* favoring natural $\underline{H}$ and H$\underline{H}$ [4,5], which invalidates [1].

(i) *Atom hydrogen-antihydrogen difference.* With relatively accurate Bohr theory, energies of left- and right-handed atoms are *degenerate* since *pure Coulomb forces* are identical for $e_1e_2$ (H) and $e_2e_1$ ($\underline{H}$). For Bohr, a distinction is *purely conventional*, meaning that Bohr theory is *achiral*. It is then normal to interpret small errors of *achiral* Bohr theory as *signatures for chiral behavior*, a very simple but overlooked solution. Yet in sophisticated bound state QED, *errors of achiral Bohr theory are explained with a quartic, which is very suspicious* as a quartic for a neutral 2-fermion system points to its *chiral behavior* [4]. *This observed quartic proves that stable $\underline{H}$-states exist* [4], contradicting the basis of [1].

(ii) *Hydrogen-antihydrogen interaction.* Pure Coulomb effects on 4-fermion system stability must be assessed unambiguously before validating [1]. The HH non-relativistic 10 term Hamiltonian **H**$_+$ = **H**$_0$+Δ**H** has atomic threshold **H**$_0$ and perturbation +Δ**H**, consisting of *4 pure Coulomb terms*. Then, H$\underline{H}$ *charge-conjugated* Hamiltonian **H**$_-$ = **H**$_0$-Δ**H** would suggest *without proof* that *charge-anti-symmetrical* H$\underline{H}$-states are repulsive, in line with [1], iff *charge-symmetrical* HH-states give *stable* H$_2$. Mutually exclusive **H**$_\pm$ =**H**$_0$±Δ**H** contradict the Heitler-London convention that stable H$_2$ is *charge-symmetrical* HH, since it can be *proved* theoretically and experimentally [5a] that stable H$_2$ is *charge-anti-symmetrical* H$\underline{H}$. Errors with H and H$_2$ symmetries contradict proof [1], *as both $\underline{H}$ and H$\underline{H}$ exist in nature and are stable* [4-5]. These arguments suffice to flaw [1] but *pure Coulomb effects for H$\underline{H}$* have even more direct implications [5b].

(iii) *Hydrogen-antihydrogen oscillations* [6]. The energy difference δ between states HH and H$\underline{H}$ in (ii) is

δ= **H**$_0$-Δ**H**–(**H**$_0$ +Δ**H**)=-2Δ**H**    (1)

a *pure Coulomb effect*, involving $\underline{H}$. To make sense, H-$\underline{H}$ oscillations hν must obey *pure Coulomb quantum gap* δ, iff hν=δ. Scaling gap δ gives

δ'=δ/($e^2/r_0$)=-2$r_0$(-1/$r_{bA}$–1/$r_{aB}$+1/$r_{ab}$+1/$r_{AB}$)  (2).

With $r_{AB}$=R, $r_{aA}$=$r_{bB}$ =$r_0$=0,5291 Å and with the 2 leptons rotating in phase in planes, perpendicular to R, *the pure Coulomb dipole-dipole effect* gives

δ'=δ/($e^2/r_0$)=-4(0,5291/R)[1-(1+(0,5291/R)$^2$)$^{-½}$] (3)

*a genuine ab initio theoretical result for pure Coulomb long-range effects, with the prospect of detecting H-$\underline{H}$-oscillations.*

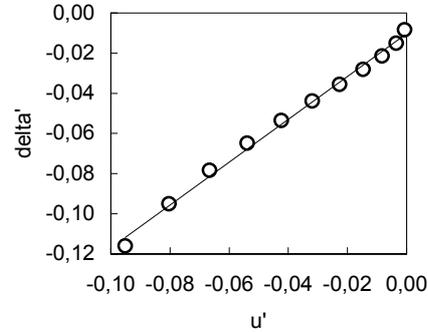

Fig. 1 δ' vs u'

The H$_2$ potential energy curve [7] gives *observed long-range behavior*, with energies u' =(U$_\infty$-U$_R$)/($e^2/r_0$) for 11 outer turning points below the threshold. The linear plot of δ' versus u' in Fig. 1

δ'=1,0667u'-0,0103 (fit R$^2$=0,9945)    (4)

*is an ab initio proof that H-$\underline{H}$ oscillations occur in nature* [5b]. *Pure Coulomb effect* (3) *for stable H$\underline{H}$*, completely neglected in [1], even solves the mystery with H-$\underline{H}$ oscillations (and B-L symmetry breaking) [6]. *With unstable $\underline{H}$*, H-$\underline{H}$ oscillation times are 10$^{20}$ s in the SM [6] (as in [1]). *With stable $\underline{H}$* (as in [4,5]), these are 10$^{-15}$ s, *a common sense but large discrepancy of* 10$^{35}$!

(iv) *Matter-antimatter asymmetry* [9]. *The pure Coulomb results* (ii-iii) *probing stable H$\underline{H}$ but unjustly disregarded in* [1], can even solve this cosmological problem [9]. The quartic in (i) proves that matter H is different from antimatter $\underline{H}$ [4]. But with (ii)-(iii) it is evident that amounts of matter H and antimatter $\underline{H}$ in *stable* H$\underline{H}$ (H$_2$) *must be equal for classical stochiometric reasons. Hydrogen being the most abundant species in the Universe, this long-standing difficult problem is simply removed* [5].

*We falsify claim* [1], *inspired by* [10-11]*, since Coulomb effects (i-iv) prove that 2- and 4-fermion systems $\underline{H}$ and H$\underline{H}$ are natural and stable* [4,5]*, instead of unstable* [1].